\documentstyle[12pt]{article}
\def\lsim{\mathrel{\rlap {\raise.5ex\hbox{$ < $}}
{\lower.5ex\hbox{$\sim$}}}}
\def\gsim{\mathrel{\rlap {\raise.5ex\hbox{$ > $}}
{\lower.5ex\hbox{$\sim$}}}}
\newcommand{\vsone}{\vspace{1cm}}
\newcommand{\pr}{\paragraph{}}
\newcommand{\be}{\begin{equation}}
\newcommand{\ee}{\end{equation}}
\newcommand{\bea}{\begin{eqnarray}}
\newcommand{\nn}{\nonumber}
\newcommand{\eea}{\end{eqnarray}}
\newcommand{\nd}[1]{/\hspace{-0.6em} #1}
\newcommand{\nk}{\noindent}

\def\gappeq{\mathrel{\rlap {\raise.5ex\hbox{$>$}}
{\lower.5ex\hbox{$\sim$}}}}

\def\lappeq{\mathrel{\rlap{\raise.5ex\hbox{$<$}}
{\lower.5ex\hbox{$\sim$}}}}

\begin{document}
\begin{titlepage}
\begin{flushright}
CERN-TH./95-297  \\
OUTP-95-43P\\
hep-th/9511071\\
\end{flushright}

\begin{centering}
\vspace{.1in}
{\large {\bf Dilatonic Black Holes \\ \vspace{.2in}
in Higher Curvature String Gravity}} \\
\vspace{.4in}
{\bf P. Kanti }$^{(a)}$,
{\bf N.E. Mavromatos }$^{(b)\,\,*}$, {\bf J. Rizos }$^{(c)}$,\\
{\bf K. Tamvakis }$^{(c)\,\,(d)}$ and {\bf E. Winstanley }$^{(b)}$\\
\vspace{.2in}
$^{(a)}$Division of Theoretical Physics, Physics Department,\\
  University of Ioannina, Ioannina GR-451 10, GREECE\\[3mm]
$^{(b)}$Department of Physics
(Theoretical Physics), University of Oxford,\\ 1 Keble Road,
Oxford OX1 3NP, U.K.  \\[3mm]
$^{(c)}$CERN, Theory Division, 1211 Geneva 23, SWITZERLAND\\

\vspace{.2in}
{\bf Abstract} \\
\vspace{.05in}
\end{centering}
{\small We give analytical arguments and demonstrate
numerically the existence of black hole solutions
of the $4D$ Effective Superstring Action in the
presence of Gauss-Bonnet quadratic curvature terms.\,\,The
solutions possess non-trivial dilaton hair.\,\,The hair, however,
is of ``secondary type", in the sense that the dilaton charge is
expressed in terms of the black hole mass.\,\,Our solutions
are not covered by the assumptions of existing proofs of
the ``no-hair" theorem.\,\,We also find some alternative solutions
with singular metric behaviour, but finite energy.\,\,The absence
of naked singularities in this system is pointed out.}

\vspace{0.2in}
%
\pr

\vspace{0.01in}
\begin{flushleft}
$^{(d)}$ on leave of absence from Physics Department, University
of Ioannina\\ [3mm]
$^{*}$P.P.A.R.C. Advanced Fellow\\ \vspace{0.2in}
\,\,\,November 1995 \\
\end{flushleft}
\end{titlepage}
\newpage
\section{Introduction}
\pr
It has become evident in recent years that the properties of black
holes are modified when the theory of matter fields has sufficient
structure. In the presence of the low energy degrees of freedom
characteristic of String Theory~\cite{witten}, i.e. dilatons, axions and
Abelian or Yang-Mills fields, it is possible to have non-trivial
static configurations for these fields outside the horizon,
i.e. have black holes with hair \cite{shapere} \cite{bizon}.
It is not clear however whether these cases, in which
the ``no-hair theorem" \cite{bek} does not apply \cite{mw}, represent
stable solutions.
Explicit black hole solutions have
been found also in string-effective theories involving
higher-order curvature corrections to the Einstein
gravity. They exhibit secondary hair of the dilaton,
axion and modulus fields. The solutions
were approximate, in the sense that only a perturbative
analysis to ${\cal O}(\alpha ')$~\cite{kanti} and
${\cal O}(\alpha^{'2})$ \cite{mignemi} has been performed.
This analysis motivates the search for
exact (to all orders in $\alpha '$) solutions
within the framework of
curvature-squared corrections to Einstein's theory.
Although the effect of the higher-order curvature terms
is not small for energy scales of order $\alpha '$,
 from a local field theory point of view it makes sense to
look for this kind of solutions, with the hope of
drawing some useful conclusions that might be of relevance
to the low-energy limit of string theories.
\pr
In the present article we shall demonstrate the existence of black hole
solutions of the Einstein-dilaton system in the presence of the
higher-derivative, curvature squared terms. These solutions will be
endowed with a non-trivial dilaton field outside the horizon, thus
possessing dilaton hair. The treatment of the quadratic terms will be
non-perturbative and the solutions are present for any value of
$\alpha'/g^2$.
What we shall argue in this paper is that the presence of these terms
provides the necessary `repulsion' in the effective
theory that balances the gravitational
attraction, thereby leading to
black holes dressed with non-trivial classical
dilaton hair.
An analogous phenomenon occurs already in the case
of Einstein-Yang-Mills systems\cite{bizon}. There, the presence of
the non-abelian gauge field repulsion balances the
gravitational attraction leading to black hole
solutions with non-trivial gauge and scalar (in Higgs systems) hair.
\pr
It is useful to discuss briefly the
situation in effective theories obtained from the string.
We shall concentrate on the bosonic part of the
gravitational multiplet
which consists of the dilaton, graviton,
and antisymmetric tensor fields. In this work we shall ignore
the antisymmetric tensor for simplicity\footnote{In four dimensions,
the antisymmetric tensor field leads to the axion hair, already
discussed in ref.\cite{kanti}. Modulo unexpected surprises,
we do not envisage problems
associated with its presence as regards
the results discussed in this work, and, hence,
we ignore it for simplicity.}.
 As is well known in low-energy effective field theory,
there are ambiguities in the coefficients of such terms,
due to the possibility of {\it local} field redefinitions which
leave the $S$-matrix amplitudes of the effective
field theory invariant,
according to the {\it equivalence} theorem.
To ${\cal O}(\alpha ')$ the freedom of such redefinitions
is restricted to two generic structures, which cannot be removed by
further redefinitions~\cite{metsaev}. One is a
 curvature-squared combination,
and the other is a four-derivative dilaton term.
Thus, a generic form of the string-inspired  ${\cal O}(\alpha ')$
corrections to Einstein's gravitation have the form
\be
{\cal L}=-\frac{1}{2}R - \frac{1}{4} (\partial _\mu \phi )^2
+ \frac{\alpha '}{8g^2} e^{\phi } (c_1{\cal R}^2  + c_2
(\partial _\rho \phi )^4 )
\label{one}
\ee
where $\alpha '$ is the Regge slope, $g^2$
is some gauge coupling constant (in the case
of the heterotic string that we concentrate
for physical reasons), and
${\cal R}^2$ is a generic curvature-dependent
quadratic  structure, which can always be fixed
to correspond to the Gauss-Bonnet (GB) invariant
\be
R^2_{GB} =
R_{\mu\nu\rho\sigma}R^{\mu\nu\rho\sigma} -
4 R_{\mu\nu}R^{\mu\nu} + R^2
\label{two}
\ee
The coefficients $c_1$, $c_2$ are fixed by comparison
with string scattering amplitude computations,
or $\sigma$-model $\beta$-function analysis. It is known that
in the three types of string theories, Bosonic, Closed-Type II Superstring,
and Heterotic Strings, the ratio of the $c_1$ coefficients
is 2:0:1 respectively~\cite{metsaev}.
The case of superstring II effective theory, then,
is characterized by the absence of curvature-squared terms.
In such theories the fourth-order dilaton terms can still be, and in
fact they are, present. In such a case, it is straightforward to
see from the modern proof of the no-scalar hair theorem
of ref. \cite{bek} that such theories, cannot sustain to
order ${\cal O}(\alpha ')$, any non-trivial dilaton hair.
On the other hand, the presence of curvature-squared terms
can drastically change the situation,
as we shall discuss in this article.
There is a simple reason to expect that in this case
the no-scalar-hair theorem  can be bypassed.
In the presence of curvature-squared terms,
the modified Einstein's equation leads to an effective
stress tensor that involves the gravitational field.
This implies that the assumption of
positive definiteness of the time-component of this tensor,
which in the Einstein case is the local energy density
of the field, may,
and as we shall show it does indeed,  break down.
The second, but equally important, reason, is that
as a result of the higher-curvature terms, there is an induced
modification of the relation $T_t^t = T_\theta^\theta$
between the time and angular components of the stress tensor,
which was valid
in the case of
spherically-symmetric Einstein theories of ref. \cite{bek}.
\pr
The structure of the article is the following: In section 2 we give
analytic arguments for the existence of scalar (dilaton) hair
of the black hole solution, which bypasses the conditions for the
no-hair theorem. In section 3 we present an analysis of the
black hole solutions. In section 4 we discuss alternative solutions,
some of which are interesting due to the finite energy-momentum tensor
they possess. Finally,
conclusions and outlook are presented in section 5.

\section{Existence of Hair in Gravity with a Gauss-Bonnet term:
analytic arguments}
\pr
Following the above discussion we shall ignore, for simplicity,
the fourth-derivative dilaton terms
in (\ref{one}), setting from now on $c_2=0$.
However, we must always bear in mind that such terms are non-zero
in realistic effective string cases, once the GB combination
is fixed for the gravitational ${\cal O}(\alpha ')$ parts.
Then, the lagrangian for dilaton gravity with a
Gauss Bonnet term reads
\be
{\cal L}=-\frac{1}{2}R - \frac{1}{4} (\partial _\mu \phi )^2
+ \frac{\alpha '}{8g^2} e^{\phi } R^2 _{GB}
\label{three}
\ee
where  $R^2_{GB}$ is the Gauss Bonnet (GB) term (\ref{two}).
\pr
As we mentioned in the introduction, although we view (\ref{three})
as a heterotic-string effective action, for simplicity, in this paper
we shall ignore the modulus and axion
fields, assuming reality of the dilaton ($S=e^{i\phi}$ in
the notation of ref. \cite{kanti}).
We commence our analysis by noting that
the dilaton field and Einstein's equations derived from (\ref{three}),
are
\bea
&~&\frac{1}{\sqrt{-g}} \partial _\mu [\sqrt{-g} \partial ^\mu \phi ]
=-\frac{\alpha '}{4g^2} e^\phi R^2_{GB}
\label{foura} \\[3mm]
&~&R_{\mu\nu} - \frac{1}{2} g_{\mu\nu} R  =
- \frac{1}{2} \partial _\mu \phi
\partial _\nu \phi + \frac{1}{4} g_{\mu\nu} (\partial _\rho \phi )^2  -
\alpha ' {\cal K}_{\mu\nu}
\label{fourb}
\eea
where
\be
{\cal K}_{\mu\nu}=(g_{\mu\rho}g_{\nu\lambda}+g_{\mu\lambda}g_{\nu\rho})
\eta^{\kappa\lambda\alpha\beta} D _\gamma
[{\tilde R}^{\rho\gamma}_{\,\,\,\,\,\alpha\beta} \partial _\kappa f]
\ee
and
\bea
\eta ^{\mu\nu\rho\sigma} &=& \epsilon ^{\mu\nu\rho\sigma}
(-g)^{-\frac{1}{2}}\nn \\[2mm]
\epsilon ^{0ijk} &=& -\epsilon_{ijk} \nn \\[2mm]
{\tilde R}^{\mu\nu}_{\,\,\,\,\,\kappa\lambda} &=& \eta^{\mu\nu\rho\sigma}
R_{\rho\sigma\kappa\lambda}  \\[2mm]
f & = & \frac {e^\phi} {8 g^2}\nn
\eea
{}From the right-hand-side
of the modified Einstein's equation (\ref{fourb}),
one can construct a conserved
``energy momentum tensor'', $\nabla _\mu T^{\mu\nu} = 0$,
\bea
T_{\mu\nu} =  \frac{1}{2} \partial_\mu\phi \partial_\nu \phi - \frac{1}{4}
g_{\mu\nu} (\partial _\rho \phi )^2  + \alpha ' {\cal K}_{\mu\nu}
\label{five}
\eea
It should be stressed that the time component of $-T_{\mu\nu}$,
which in Einstein's gravity would correspond to the
local energy
density ${\cal E}$, may {\it not be positive }.
Indeed, as we shall see later on,
for spherically-symmetric space times, there are regions
where this quantity is negative.
The reason is that, as a result of the higher derivative GB terms,
there are contributions of the gravitational field itself
to $T_{\mu\nu}$. From a string theory point of view, this is
reflected in the fact that the dilaton is part of the string
gravitational multiplet.
Thus, this is the {\it first} important indication
on the possibility of evading the no-scalar-hair theorem
of ref. \cite{bek} in this case.
However, this by itself is not
sufficient for a rigorous proof of an
evasion of the no-hair conjecture.
We shall come to this point later on.
\pr
At the moment, let us
consider a spherically symmetric space-time having the metric
\be
ds^2 = -e^\Gamma dt^2 + e^\Lambda dr^2 + r^2 (d\theta ^2 + sin^2 \theta
d\varphi^2)
\label{six}
\ee
where $\Gamma$, $\Lambda$ depend on $r$ solely.
Before we proceed to study the above system it is useful to note that
if we turn off the Gauss-Bonnet term, equation (\ref{foura}) can be
integrated to give $\phi'\sim \frac{1}{r^2} e^{(\Lambda-\Gamma)/2}$.
A black hole solution should have at the horizon $r_h$ the behaviour
$e^{-\Gamma}$, $e^\Lambda \rightarrow \infty$. Therefore the radial
derivative of the dilaton would diverge on the horizon resulting into
a divergent energy-momentum tensor
\be
T^t_t=-T^r_r= T^\theta _\theta=-\frac{e^{-\Lambda}}{4} \phi'^2
\rightarrow \infty
\ee
Rejecting this solution we are left with the standard Schwarzschild
solution and a trivial $(\phi=constant)$ dilaton, in agreement with
the no-hair theorem. This behaviour will be drastically modified
by the Gauss-Bonnet term.
\pr
The $r$ component of the energy-momentum
conservation equations reads:
\be
\partial _r [ \sqrt{-g} T_r^r ]-\frac{1}{2}
\sqrt{-g} (\partial _r g_{\alpha\beta})
T^{\alpha\beta} = 0
\label{seven}
\ee
The spherical symmetry of the space-time implies $T_\theta^\theta =
T_\varphi^\varphi $. Then eq. (\ref{seven}) becomes:
\be
(e^{(\Gamma+\Lambda)/2}r^2 T_r^r )' =\frac{1}{2} e^{(\Gamma +
\Lambda)/2} r^2
[\Gamma ' T_t^t + \Lambda' T_r^r +
\frac{4}{r}T_\theta^\theta]
\ee
where the prime denotes differentiation with respect to $r$.
It can be easily seen that the terms containing $\Lambda $ cancel
to give
\be
(e^{\Gamma /2}r^2 T_r^r )' =\frac{1}{2} e^{\Gamma /2} r^2
[\Gamma ' T_t^t +
\frac{4}{r}T_\theta^\theta]
\ee
Integrating over the radial coordinate $r$ from the horizon $r_h$
to generic $r$ yields
\be
T_r^r (r)  =  \frac{e^{-\Gamma /2}}{2r^2} \int _{r_h}^r
e^{\Gamma /2} r^2 [\Gamma ' T_t^t + \frac{4}{r}
T_\theta ^\theta ] dr
\label{eight}
\ee
The boundary terms on the horizon vanish, since scalar invariants
such as $T_{\alpha\beta}T^{\alpha\beta}$ are finite there.
For the first derivative of $T_r^r$ we have
\bea
(T_r^r)'(r) &=& \frac{\Gamma '}{2} T_t^t + \frac{2}{r} T_\theta ^\theta
- \frac{e^{-\Gamma /2}}{r^2} (e^{\Gamma /2} r^2 )' T_r^r \nn \\ [3mm]
&=& \frac {\Gamma'} {2} (T^t_t-T^r_r) + \frac {2} {r} (T^\theta_\theta-
T^r_r)
\label{nine}
\eea
Taking into account (\ref{five}) and (\ref{six}), one easily obtains
\bea
T^t_t&=&- e^{-\Lambda} \frac {\phi'^2}{4}-
\frac {\alpha'}{g^2 r^2}e^{\phi-\Lambda} (\phi''+\phi'^2)
 (1-e^{-\Lambda}) + \frac {\alpha'} {2 g^2 r^2} e^{\phi-\Lambda}
 \phi' \Lambda' (1-3 e^{-\Lambda}) \nn \\ [3mm]
T_r^r&=& e^{-\Lambda} \frac {\phi'^2} {4}-
\frac {\alpha'} {2 g^2 r^2} e^{\phi-\Lambda} \phi'
 \Gamma' (1-3 e^{-\Lambda})  \\ [3mm]
T_\theta^\theta&=&- e^{-\Lambda} \frac {\phi'^2}{4}
 +\frac {\alpha '}{2 g^2 r} e^{\phi-2 \Lambda}
 [\Gamma'' \phi' + \Gamma' (\phi'' + \phi'^2) + \frac {\Gamma'
 \phi'} {2} (\Gamma'-3 \Lambda')] \nn
\label{comp}
\eea
In the relations (16) there lies the {\it second} reason
for a possibility of an evasion of the no-hair conjecture.
Due to the presence
of the higher curvature contributions
, the relation $T_t^t = T_\theta^\theta$
assumed in ref. \cite{bek}, is no longer valid.
The alert reader must have noticed, then,
the similarity of the r\^ole played by the Gauss-Bonnet
${\cal O}(\alpha ')$ terms in the lagrangian (\ref{three})
with the case of the non-Abelian gauge black holes
studied in ref. \cite{mw}. There,
the presence of the non-abelian gauge field repulsive forces also
lead to
non-trivial contributions to $T_\theta^\theta \ne T_t^t$,
leading to a sort
of `balancing' between this repulsion and the gravitational
attraction.
We stress once, again, however, that in our case
{\it both} the non-positivity of the ``energy-density'' $T_t^t$,
and the modified relation $T_t^t \ne T_\theta^\theta$,
play equally important r\^oles in leading to a possibility
of having non-trivial classical scalar (dilaton)
hair in GB black holes systems.
Below we shall demonstrate rigorously this, by showing that
there is {\it no contradiction} between the
results following from the conservation
equation of the ``energy-momentum tensor'' $T_{\mu\nu}$ and
the field equations, in the presence of non trivial dilaton hair.
\pr
One can define functionals ${\cal E}$, ${\cal J}$ and ${\cal G}$
by
\bea
{\cal E} &=&-T_t^t  \nn \\[2mm]
{\cal J} &=& T_r^r -T_t^t  \\[2mm]
{\cal G} &=& T_\theta^\theta-T_t^t\nn
\label{func}
\eea
Then, we can rewrite
(\ref{eight}), (\ref{nine}) in the form
\bea
T_r^r (r) &=& \frac{e^{-\Gamma /2}}{r^2}\int _{r_h}^r [-(e^{\Gamma /2} r^2)'
{\cal E} + \frac{2}{r} {\cal G} ] dr \nn \\[3mm]
(T_r^r)'(r) &=& -\frac{e^{-\Gamma /2}}{r^2} (e^{\Gamma /2} r^2)' {\cal J}
+ \frac{2}{r} {\cal G}
\label{twelve}
\eea
\pr
{}From the Einstein's equations of this system
we have
\be
e^{-\Lambda} [\frac{1}{r^2} - \frac{\Lambda '}{r} ] -\frac{1}{r^2}
= T_t^t = -{\cal E}
\ee
which can be integrated to give
\be
e^{-\Lambda}=1 - \frac{1}{r} \int _{r_h}^r {\cal E} r^2 dr -
\frac{2}{r} {\cal M}_0
\ee
where ${\cal M}_0$ is a constant of integration.
In order that $e^\Lambda  \rightarrow \infty $ as $r \rightarrow r_h$,
${\cal M}_0$ is fixed by
\be
{\cal M}_0 = \frac{r_h}{2}
\ee
Far away from the origin the unknown functions $\phi(r)$, $e^{\Lambda(r)}$,
and $e^{\Gamma(r)}$ can be expanded in a power series in $1/r$. These
expansions, substituted back into the equations, are finally expressed
in terms of three parameters only, chosen to be $\phi_{\infty}$,
the asymptotic value of the dilaton, the ADM mass $M$, and the dilaton charge
$D$ defined as~\cite{mitra}
\be
D=-\frac{1}{4\pi}\int d^2\Sigma^\mu \nabla_\mu \phi
\ee
where the integral is over a two-sphere at spatial infinity.
The asymptotic solutions are
\bea
e^{\Lambda(r)}&=&1+\frac{2 M}{r}+ \frac{4 M^2-D^2}{r^2}
+{\cal O}(1/r^3) \\ [3mm]
e^{\Gamma(r)}&=&1-\frac{2 M}{r} + {\cal O}(1/r^3) \\ [3mm]
\phi(r)&=& \phi_{\infty}+\frac{D}{r}-\frac{2 M D}{r^2}
+ {\cal O}(1/r^3)
\label{thirteen}
\eea
To check the possibility of the evasion of the no-hair conjecture
we first consider the asymptotic behaviour
of $T_r^r$ as $r \rightarrow \infty$.
In this limit, $e^{\Gamma/2} \rightarrow 1$, and so the leading behaviour
of $(T_r^r)'$ is
\be
  (T_r^r)'  \sim \frac{2}{r} [{\cal G} - {\cal J} ]=
\frac{2}{r} (T^\theta_\theta - T^r_r)
\ee
Since $\Gamma '$ and $\Lambda '$
$\sim {\cal O}(\frac{1}{r^2}) $ as $r \rightarrow \infty$,
we have the following asymptotic behaviour
\bea
  T^\theta_\theta &\sim &-\frac{1}{4} (\phi ')^2 + {\cal O}(\frac{1}{r^6})
\nn \\ [3mm]
T^r_r & \sim & \frac{1}{4} (\phi ')^2 + {\cal O}(\frac{1}{r^6})
\eea
Hence, the integral defining $T_r^r$ converges and
\be
 (T_r^r)' \sim -\frac{1}{r} (\phi ')^2 < 0 \qquad {\rm as~r~\rightarrow
\infty}
\ee
Thus, $T_r^r$ is positive and decreasing as $r \rightarrow \infty$.
\pr
We now turn to the behaviour of the unknown functions at the event horizon.
When $r \sim r_h$, we make the ansatz
\bea
e^{-\Lambda(r)}&=&\lambda_1 (r-r_h) + \lambda_2 (r-r_h)^2 +...\nn \\[3mm]
e^{\Gamma(r)}&=&\gamma_1 (r-r_h) + \gamma_2 (r-r_h)^2 +...  \\[3mm]
\phi(r)&=& \phi_h + \phi'_h (r-r_h) + \phi''_h(r-r_h)^2 +...\nn
\label{fourteen}
\eea
with the subscript $h$ denoting the value of the
respective quantities at the horizon. The consistency of (29)
will be checked explicitly in the next section.
As we can see, $\phi(r_h)\sim constant$ while $\Gamma '$ and $\Lambda '$
diverge as $(r-r_h)^{-1}$ and $-(r-r_h)^{-1}$ respectively. Then, the
behaviour of the components of
the energy-momentum tensor near the horizon is
\bea
T_r^r & = & -\frac {\alpha'} {2 g^2 r^2} e^{\phi-\Lambda} \phi'
\Gamma' + {\cal O}(r-r_h) \nn \\ [3mm]
T_t^t & = & \frac {\alpha'} {2 g^2 r^2} e^{\phi-\Lambda} \phi'
\Lambda' + {\cal O}(r-r_h)  \\ [3mm]
T_\theta^\theta &=& \frac {\alpha'} {2 g^2 r} [\Gamma'' \phi'+
\frac {\Gamma' \phi'} {2} (\Gamma'-3 \Lambda')] +
{\cal O}(r-r_h) \nn
\label{fifteen}
\eea
Taking into account the above expressions the leading behaviour
of $T_r^r$ near the horizon is
\bea
   T_r^r (r) &\simeq &\frac{e^{-\Gamma /2}}{r^2}
\int _{r_h}^r (e^{\Gamma /2})' \frac{\alpha '}{2 g^2}
e^{\phi-\Lambda} \phi' \Lambda' dr + {\cal O}(r-r_h) \nn \\[3mm]
   &=& -\frac{e^{-\Gamma /2}}{r^2} \int _{r_h}^r
\frac{\alpha '}{4g^2}e^{\Gamma/2} (\Gamma ')^2  e^{-\Lambda} e^\phi
\phi ' dr + {\cal O}(r-r_h)
\eea
Therefore one observes that
for $r$ sufficiently close to the event horizon, $T_r^r$
has {\it opposite} sign to $\phi '$.
\pr
For $(T_r^r)'$ near the horizon,
we have
\bea
-\frac{\Gamma '}{2} {\cal J} + \frac{2}{r} ({\cal G}-{\cal J})
&=& \frac{\alpha '}{2g^2} \frac{e^\phi}{r^2} e^{-\Lambda}
\{ -\Gamma ' (\phi '' + \phi'^2) +
\phi ' [ \frac{\Gamma '}{2} (\Gamma ' + \Lambda ') +
2 e^{-\Lambda} \Gamma ''\nn \\[3mm]
&~&- \frac{2}{r} \Lambda ' ] \}
-\frac{1}{4} \Gamma ' e^{-\Lambda} \phi'^2+ {\cal O}(r-r_h)
\label{fifteena}
\eea
where $\Gamma ' + \Lambda ' \sim {\cal O}(1)$ for $r \sim r_h$.
\pr
In order to simplify the above expression further, we turn to the
field equations.
Using the static, spherically symmetric ansatz (\ref{six}) for the metric,
the dilaton equation as well as the $(tt)$, $(rr)$ and $(\theta\theta)$
component of the Einstein's equations take the form
\bea
&~& \phi''+\phi'(\frac{\Gamma'-\Lambda'}{2}+\frac{2}{r})=
\frac{\alpha'e^\phi}{g^2 r^2}\left(\Gamma'\Lambda'e^{-\Lambda}+
(1-e^{-\Lambda})[\Gamma''+
\frac{\Gamma'}{2}(\Gamma'-\Lambda')]\right)
\label{sixteen} \\[3mm]
&~& \Lambda'\left(1+\frac {\alpha' e^\phi} {2 g^2 r} \phi'
(1-3 e^{-\Lambda})\right)
=\frac{r\phi'^2}{4}+\frac {1-e^\Lambda}{r}
+\frac{\alpha' e^\phi}{g^2 r}(\phi''+\phi'^2)(1-e^{-\Lambda})
\label{seventeen} \\ [3mm]
&~& \Gamma'\left(1+\frac {\alpha' e^\phi}{2 g^2 r}\phi'
(1-3 e^{-\Lambda})\right)
=\frac{r \phi'^2}{4}+\frac{e^\Lambda-1}{r}
\label{eighteen} \\[3mm]
&~& \Gamma''+\frac{\Gamma'}{2}(\Gamma'-\Lambda')+
\frac{\Gamma'-\Lambda'}{r}
=-\frac{{\phi'}^2}{2}+\frac{\alpha' e^{\phi-\Lambda}}{g^2 r}
\left(\phi'\Gamma''+(\phi''+{\phi'}^2)\Gamma'\right.\nn \\[3mm]
&~&\hspace{5.4cm}\left. +\frac{\phi'\Gamma'}{2}(\Gamma'-3\Lambda')\right)
\label{nineteen}
\eea
At the event horizon $r \sim r_h$ the $(tt)$ and $(rr)$
components reduce to
\bea
e^{-\Lambda} \Lambda ' &=& -\frac{1}{r} - \frac{\alpha '}{2g^2}
\frac{e^\phi}{r} e^{-\Lambda} \Lambda ' \phi ' + {\cal O}(r-r_h)
\nn \\[3mm]
e^{-\Lambda} \Gamma ' &=& \frac{1}{r} -\frac{ \alpha '}{2g^2}
\frac{e^\phi}{r} e^{-\Lambda} \Gamma ' \phi '
+ {\cal O}(r-r_h)
\label{twenty}
\eea
Hence, $e^{-\Lambda} \Lambda ' = - \frac{1}{{\cal F} r} + {\cal O}(r-r_h)$,
$e^{-\Lambda} \Gamma ' = \frac{1}{{\cal F} r} + {\cal O}(r-r_h) $,
with
\be
{\cal F} = 1 + \frac{\alpha '}{2g^2} \frac{e^{\phi _h}}{r_h}
\phi'_h
\label{twentya}
\ee
{}From the $(\theta\theta)$ component of the Einstein's equations
we obtain
\bea
e^{-2\Lambda} \Gamma '' &=& -\frac{1}{2} e^{-2\Lambda}
(\Gamma ')^2 + \frac{1}{2} e^{-2\Lambda} \Gamma ' \Lambda '
+ {\cal O}(r-r_h)  \nn \\[2mm]
&=&-\frac{1}{r_h^2 {\cal F}^2 } + {\cal O}(r-r_h)
\label{twentyb}
\eea
Finally, adding the $(tt)$ and $(rr)$ components we obtain
\be
\Gamma ' + \Lambda ' = \frac{1}{{\cal F}} [ \frac{1}{2}
r_h (\phi _h ')^2 + \frac{\alpha '}{g^2} \frac{e^{\phi _h}}{r_h}
(\phi _h'' + (\phi _h')^2 ) ] + {\cal O}(r-r_h)
\ee
Substituting all the above formulae into (\ref{fifteena})
yields, near $r_h$
\be
  (T_r^r)' (r) \sim -\frac{1}{4}
\frac{(\phi _h')^2}{r_h^2 {\cal F}}
- \frac{\alpha '}{2g^2} \frac{e^{\phi _h}}{r_h^3 {\cal F}^2 }
(\phi _h'' + (\phi _h ')^2) - \frac{\alpha '}{4g^4}
\frac{e^{2\phi _h}}{r^5_h {\cal F}^2}
(\phi _h ')^2 + {\cal O}(r-r_h)
\label{twentyc}
\ee
\pr
Next, we turn to the dilaton equation (\ref{sixteen}). Combining
eqs.(\ref{twenty}) and (\ref{twentyb})
we obtain
\be
e^{-\Lambda} \{ \Gamma '' + \frac{\Gamma '}{2}
(\Gamma ' - \Lambda ') \} =
-\frac{2}{r_h^2 {\cal F}^2} + {\cal O}(r-r_h)
\ee
Then, the dilaton equation (\ref{sixteen}) at $r \sim r_h$ takes the form
\be
   \frac{\phi _h'}{r_h {\cal F}} = - \frac{3}{{\cal F}^2}
\frac{\alpha '}{g^2} \frac{e^{\phi _h}}{r_h^4} + {\cal O}(r-r_h)
\ee
from which it follows that
\be
  \phi _h ' = -\frac{3}{r_h^3 {\cal F}} \frac{\alpha '}{g^2}
e^\phi
\ee
Substituting for ${\cal F}$ (\ref{twentya}), the following equation
for $\phi_h '$ is derived
\be
 \frac{\alpha '}{2g^2} \frac{e^{\phi _h}}{r_h}
(\phi _h ')^2+ \phi'_h+ \frac{3}{r_h^3} \frac{\alpha '}{g^2}
e^{\phi _h} = 0
\ee
which has as solutions
\be
\phi _h' = \frac{g^2}{\alpha '}r_h e^{-\phi _h}\left(
-1 \pm \sqrt{1 - \frac{6(\alpha ')^2  }{g^4}
\frac{e^{2\phi _h}}{r_h^4}}\right)
\label{twentyfive}
\ee
As we will see below the relation (\ref{twentyfive}) guarantees
the {\it finiteness}
of $\phi _h ''$, and hence of the ``local density'' $T_t^t$ (16).
Both these solutions for $\phi_h'$ are negative, and hence, since
$T_r^r(r_h)$  has the opposite sign to $\phi _h'$, $T_r^r$
will be {\it positive} sufficiently close to the horizon.
Since $T_r^r \ge 0$ also at infinity, we observe that there is
{\it no contradiction} with Einstein's equations, thereby
allowing for the existence of black holes with scalar hair.
We observe
that near the horizon the quantity
${\cal E}$ ($-T_t^t$) (17), which in Einstein's
gravitation would be the local energy density
of the field $\phi$, is
{\it negative}. As we mentioned earlier, this
constitutes one of the reasons one should expect
an evasion of the no-scalar-hair conjecture
in this black hole space time.
Crucial also for this result was the presence of
additional terms in (16),
leading to $T_t^t \ne T_\theta^\theta$.  Both of these
features, whose absence in
the case of Einstein-scalar gravity
was
crucial for the modern proof of
the no-hair theorem,
owe their existence in
the presence
of the higher-order ${\cal O}(\alpha ')$ corrections
in (\ref{three}).
\pr
The physical importance of the restriction (\ref{twentyfive})
lies on the fact that according to this relation,
black hole solutions of a given horizon radius
can {\it only exist} if the coupling constant of the
Gauss-Bonnet term in (\ref{three}) is smaller than
a critical value, set by the magnitude of
the horizon scale. In fact from (\ref{twentyfive}), reality
of $\phi _h'$ is guaranteed if and only if
\be
    e^{\phi _h} < \frac{g^2}{\sqrt{6}\alpha '} r_h^2
\label{tsix}
\ee
By a conformal rescaling of the dilaton field we can always
set $\alpha ' /g^2 \rightarrow 1$, in which case $\phi _h$
can be viewed as a constant added to the dilaton field.
In this picture, $\beta \equiv \frac{1}{4}e^{\phi _h}$ can
then be viewed as
the (appropriately normalized with respect to the Einstein term)
coupling constant of the GB term in the effective lagrangian
(\ref{three}). For a black hole of unit horizon radius $r_h =1$,
the critical value of $\beta$, above which black hole solutions
cannot exist, is then $\beta _c = 4/\sqrt{6}$.
One is tempted to compare the situation with the case of $SU(2)$
sphaleron solutions in the presence of Gauss Bonnet
terms \cite{donets}.
Numerical analysis of sphaleron solutions in such systems reveals
the existence of a critical value for the GB coefficient
above which solutions do not exist. In the sphaleron case
this number depends on the number of nodes of the
Yang-Mills gauge field. In our case, if
one fixes the position
of the horizon, then it seems that in order to
construct black hole solutions with this horizon size
the GB coefficient has to satisfy (\ref{tsix}).
The difference of (\ref{tsix}) from the result of ref. \cite{donets}
lies on the existence of an extra scale, as compared to the sphaleron
case, that of the black hole horizon $r_h$.
Thus, the most correct way of interpreting (\ref{twentyfive}) is that
of viewing it as
providing a necessary condition for the {\it absence of naked
singularities} in space-time.
To understand better this latter point, we have to
discuss the simpler case of a dilatonic black hole
in the presence of an Abelian gauge field in four dimensions
\cite{mitra}. Such black holes
admit dilatonic non-constant hair outside their horizon
only in the presence of gauge fields conformally coupled
to the dilaton
\be
    S \propto \int d^4x \sqrt{-g} [-R - \frac{1}{2} (\partial _\rho \phi)^2
- \frac{1}{2} e^{\phi} g^{\mu\lambda} g^{\nu\rho} F_{\mu\nu}F_{\lambda\rho}]
\label{tseven}
\ee
where $F_{\mu\nu}$ is the field strength of the (Abelian) gauge field.
The metric space-time strongly resembles the Schwarzschild solution,
with an horizon $r_h =2M$, with $M$ the mass of the black hole,
and a curvature singularity at $r=a=\frac{Q^2}{2M}e^{-\phi _0}$,
where $\phi _0$ is an arbitrary constant added to the dilaton,
reflecting the conformal coupling, and
$Q$ is the magnetic charge of the black hole, associated
with $F_{\theta\phi}=Qsin\theta$. If $Q \ne 0$,
the black hole admits non-constant
dilaton configurations outside the horizon
\be
    e^{\phi}=e^{\phi _0}(1-\frac{a}{r})
\label{44}
\ee
{}From the above equation it becomes clear that
consistency of the dilaton solution requires
$r > a$, which is equivalent to the absence of
naked singularities, since a curvature singularity
arises at the point $r=a$ for this dilatonic black
hole.
A similar thing we conjecture as happening in our case,
where (\ref{tsix}) is interpreted as the necessary condition
for the absence of naked singularities. We conjecture
that in our black hole solution a curvature singularity
occurs at $r_0^2=\frac{\sqrt{6} \alpha '}{g^2}e^{\phi _h}$, and,
thus, due to (\ref{tsix}), $r_h > r_0$.
\pr
Above, we have argued on the possibility of
having black holes in the system (\ref{three}) that admit
non-trivial dilaton hair outside their horizon.
The key is the bypassing of the no-hair theorem~\cite{bek}, as a result of
the curvature-squared terms. In what follows we shall write down
explicit solutions of the equations of motion originating from
(\ref{three}) and provide evidence for the existence
of black hole solutions to all orders in $\alpha '$.
Unfortunately a complete analytic treatment
of these equations is not feasible, and one has to
use numerical methods. This complicates certain things,
in particular it does not allow for a clear view of
what happens inside the horizon, thereby not giving
any information on the curvature singularity structure.

\section{Black Hole Solutions}

\subsection{Analytic Solution near the Horizon}
\pr
We now turn to the behaviour of the fields near
the horizon. For calculational convienence we set
$\frac{\alpha'}{g^2}\rightarrow 1$ ,
shifting $\phi\rightarrow\phi-\log\left(\frac{\alpha'}{g^2}\right)$.
We start by observing that the $(rr)$ component can be solved
analytically to yield an expression for $e^\Lambda$
\be
e^\Lambda=\frac{-\beta+\delta\sqrt{\beta^2-4\gamma}}{2}\ ,\ \delta=\pm1
\label{thirty}
\ee
where
\bea
\beta&=&\frac{\phi^{'2} r^2}{4}-1-\Gamma'(r+\frac{e^\phi\phi'}{2})
 \nn \\
\gamma&=&\frac{3}{2}\Gamma'\phi'e^\phi
\eea
We, then eliminate $\Lambda'$ using $\frac{d}{dr}(rr)$.
Choosing two of the remaining equations (\ref{sixteen}),
(\ref{seventeen}) and (\ref{nineteen}) (only two of them are linearly
independent) we obtain the system of equations
\bea
&\phi''&=-\frac{d_1}{d}
\label{thone}
\\
&\Gamma''&=-\frac{d_2}{d}
\label{thtwo}
\eea
where
\bea
&d=&
4e^{2 \Lambda + \phi}r\left(-4 + 8e^\Lambda - 4e^{2 \Lambda} -
 4 \Gamma' r  + 4 \Gamma' e^\Lambda r+ 5{\phi'}^2 r^2 -
{\phi'}^2 e^\Lambda r^2\right)
\nn\\[2mm]
&~&+ 4 \phi' e^{\Lambda+2\phi}
   \left(6 - 12e^\Lambda + 6e^{2\Lambda} + 6 \Gamma' r -
8 \Gamma' e^\Lambda r + 2 \Gamma' e^{2\Lambda}r - 3{\phi'}^2r^2
\right. \nn\\[2mm]
&~&\left.+{\phi'}^2e^\Lambda r^2\right)
-12 \Gamma'{\phi'}^2e^{3\phi} (1 - e^\Lambda)^2
- 8{\phi'}e^{3{\Lambda}}r^4
\\[3mm]
&d_1=&
2 \Gamma'{\phi'}^3e^{3\phi}\left(9\Gamma' - 6\phi' - 6\Gamma'e^\Lambda
+12\phi'e^\Lambda + \Gamma'e^{2\Lambda}- 6\phi'e^{2\Lambda}\right)
\nn\\[2mm]
&~& +{\phi'}^2e^{\Lambda+2\phi}\left(24\phi' - 8\Gamma'e^\Lambda -
48\phi'e^\Lambda + 8\Gamma'e^{2\Lambda} + 24\phi'e^{2\Lambda} -
42{\Gamma'}^2r\right.\nn \\[2mm]
&~&\left. - 30{\phi'}^2r + 20{\Gamma'}^2e^\Lambda r
- 32\Gamma'\phi'e^\Lambda r +16{\phi'}^2e^\Lambda r
- 2{\Gamma'}^2e^{2\Lambda}r \right.\nn\\[2mm]
&~& \left.- 2{\phi'}^2e^{2\Lambda}r +3{\Gamma'}{\phi'}^2r^2
- 3 \Gamma'{\phi'}^2e^\Lambda r^2
+ 24\Gamma'\phi'r+ 8\Gamma'\phi'e^{2\Lambda}r\right)
\nn\\[2mm]
&~& +\phi'e^{2\Lambda+\phi}
\left(-24 + 48e^\Lambda - 24e^{2\Lambda} -
4\Gamma'r - 16\phi'r + 8\Gamma'e^\Lambda r \right.
\nn\\[2mm]
&~& \left.+ 32\phi'e^\Lambda r - 4\Gamma'e^{2\Lambda}r-
16\phi'e^{2\Lambda}r + 32{\Gamma'}^2r^2- 16\Gamma'\phi'r^2 +
38{\phi'}^2r^2\right.
\nn\\[2mm]
&~& \left.+ 16\Gamma'\phi'e^\Lambda r^2-6{\phi'}^2e^\Lambda r^2 -
3\Gamma'{\phi'}^2r^3 +\Gamma'{\phi'}^2e^\Lambda r^3
- 8{\Gamma'}^2e^\Lambda r^2\right)\nn \\[2mm]
&~& +2e^{3\Lambda}r\left(8 - 16e^\Lambda + 8e^{2\Lambda} + 4\Gamma'r -
4\Gamma'e^\Lambda r - 4{\Gamma'}^2r^2 - 6{\phi'}^2r^2\right.
\nn\\[2mm]
&~& \left.+ \Gamma'{\phi'}^2r^3 - 2{\phi'}^2e^\Lambda r^2\right)
\\[3mm]
&d_2=&
\Gamma'\phi'e^{\Lambda+2\phi} r \left(18{\Gamma'}^2 + 6{\phi'}^2 -
 4{\Gamma'}^2e^\Lambda + 8{\phi'}^2e^\Lambda + 2{\Gamma'}^2e^{2\Lambda} +
 2{\phi'}^2e^{2\Lambda}\right.
\nn\\[2mm]
&~&\left. + 5\Gamma'{\phi'}^2e^\Lambda r - 8{\phi'}^3e^\Lambda r
 - 9\Gamma'{\phi'}^2r \right)
-2{\Gamma'}^3{\phi'}^2e^{3\phi}\left(3 + e^{2\Lambda}\right)
\nn\\[2mm]
&~&
+\phi'e^{3\Lambda}r^2\left(8 - 8e^\Lambda - 4\Gamma'r -
4\Gamma'e^\Lambda r - 4{\Gamma'}^2r^2 - 2{\phi'}^2r^2 +
\Gamma'{\phi'}^2r^3\right)
\nn\\[2mm]
&~&
+ e^{2\Lambda+\phi}\left(8\Gamma'-16\Gamma'e^\Lambda+8\Gamma'e^{2\Lambda}
- 4{\Gamma'}^2r + 8{\phi'}^2r + 8{\Gamma'}^2e^\Lambda r\right.
\nn\\[2mm]
&~&\left.-4{\Gamma'}^2e^{2\Lambda}r-8{\phi'}^2e^{2\Lambda}r
- 12{\Gamma'}^3r^2 -
 10\Gamma'{\phi'}^2r^2 - 8{\phi'}^3r^2 + 4{\Gamma'}^3e^\Lambda r^2
\right.\nn\\[2mm]
&~& \left.+ 2\Gamma'{\phi'}^2e^\Lambda r^2+
8{\phi'}^3e^\Lambda r^2 + 13{\Gamma'}^2{\phi'}^2r^3+4\Gamma'{\phi'}^3r^3
+ 6{\phi'}^4r^3\right.
\nn\\[2mm]
&~&\left.+ 4\Gamma'{\phi'}^3e^\Lambda r^3 - 2{\phi'}^4e^\Lambda r^3 -
3\Gamma'{\phi'}^4r^4- 3{\Gamma'}^2{\phi'}^2e^{\Lambda}r^3 \right)
\eea
\pr
Assuming $\phi_h$ and $\phi_h'$ to be finite and $\Gamma'\rightarrow\infty$
when $r\rightarrow r_h$, we expand the rhs of (\ref{thirty}) near
the horizon
\be
e^\Lambda=\frac{1}{2}(e^\phi \phi'+2 r)\Gamma'-
\frac{\textstyle 8e^\phi \phi'-8r+e^\phi {\phi'}^3r^2+2 {\phi'}^2r^3}
{\textstyle 4(e^\phi \phi'+2 r)}
+{\cal O}\left({\frac{1}{\Gamma'}}\right)
\label{exple}
\ee
for $(e^\phi \phi'+2 r) \ne0$ and $\delta = 1$~\footnote{This has to be
understood by the following facts: (i) the choice $\delta=-1$ leads to
$e^{\Lambda}={\cal O}(1)$ near the horizon, which is not a black hole
solution, and (ii) if $(e^\phi \phi'+2 r)\simeq0$,
then one obtains $e^\Lambda = \sqrt{3 r} (\Gamma')^{1/2} + ...$ and
$\phi'' =  \sqrt{3 r} e^{-\phi} (\Gamma')^{1/2}+...$
which implies that $\phi''_h$ is finite at the horizon only if
$\phi_h \rightarrow \infty$. This is inconsistent with our initial
assumption for finite $\phi_h$ and $\phi'_h$.}.
Substituting (\ref{exple}) in (\ref{thone}), (\ref{thtwo}) we obtain
\bea
\phi'' &=& -\frac{\textstyle1}{\textstyle2}\frac{\textstyle (e^\phi \phi'+2
 r)(6e^\phi+e^\phi{\phi'}^2
r^2+2 \phi'r^3)}{\textstyle -6 e^{2\phi}+e^{\phi}\phi'r^3+2 r^4}\Gamma'
+{\cal O}(1)
\label{ddphi}
\\
\Gamma'' &=& -\frac{\textstyle1}{\textstyle2}\frac{\textstyle -6
 e^{2\phi}+e^{2\phi}\phi'r^2+4 e^{\phi}\phi'r^3
+4r^4}{\textstyle -6 e^{2\phi}+e^{\phi}\phi'r^3+2 r^4}{\Gamma'}^2
 +{\cal O}(\Gamma')
\label{ddn}
\eea
We now observe that in order to keep $\phi''_h$ finite we have
to impose the boundary condition
$6e^\phi+e^\phi {\phi'}^2
r^2+2 \phi'r^3=0$ which relates $\phi'_h$ with $\phi_h$
\bea
\phi'_h &=&r_h e^{-\phi_h}\left(-1+\sigma\sqrt{1-6\frac{e^{2\phi_h}}
{r_h^4}}\right) \ , \sigma=\pm1
\nn\\[3mm]
\phi_h &<& \log(\frac{r^2_h}{\sqrt{6}})
\label{conea}
\eea
and implies
\bea
\phi'' &=& {\cal O}(1)\\[2mm]
\Gamma''&=&-{\Gamma'}^2+{\cal O}(1)\Rightarrow
\Gamma' = \frac{\textstyle1}{\textstyle r-r_h}+{\cal O}(1)\nn
\eea
or
\bea
e^{\Gamma(r)} &=& \gamma_1 (r-r_h)+{\cal O}(r-r_h)^2 \nn \\[3mm]
e^{-\Lambda(r)}&=& \lambda_1 (r-r_h)+{\cal O}(r-r_h)^2
\label{coneb}
\eea
where $\gamma _1$ is an arbitrary constant, and
$\lambda _1 = 2/(e^{\phi _h}\phi _h' + 2r_h)$.
Notice that (\ref{conea}) is exactly the equation (\ref{twentyfive}),
obtained previously, from the dilaton equation of motion
near the horizon. This shows that if (\ref{conea}) is
satisfied, then not only the finiteness of $\phi _h ''$ is guaranteed
but also the expected singular behaviour of the
metric is assured. The above analysis leads to the
conclusion that the asymptotic solution (\ref{conea})-(\ref{coneb}),
is the only acceptable
black hole solution with finite $\phi_h$, $\phi'_h$ and $\phi''_h$.

\subsection{Numerical Analysis}
\pr
We now proceed to the numerical integration. Starting from the
solution (\ref{conea})-(\ref{coneb}), at $r=r_h+\epsilon$ ,
$\epsilon\simeq O(10^{-8})$,
we integrate the system (\ref{thone}), (\ref{thtwo}) towards
$r \rightarrow \infty$ using the
fourth order Runge-Kutta method with an automatic step procedure
and accuracy $10^{-8}$. The integration stops when the flat
space-time asymptotic limit (23)-(25) is reached. Since $\Lambda(r)$
is not an independent variable, and $\phi_h'$ is related to $\phi_h$
and $r_h$ through eq.(\ref{conea}), it seems that the only
independent parameters of the problem are $\phi_h$, $r_h$ and
$\gamma_1$. Note that the
equations of motion do not yield any constraint for $\gamma_1$.
This is due to the fact that
the equations of motion (\ref{sixteen})-(\ref{nineteen}) do not
involve $\Gamma(r)$ but only $\Gamma'(r)$. Thus, only $\Gamma'(r)$ can be
determined by them and in order to obtain $\Gamma(r)$ a final integration
has to be performed. This integration involves an integration constant
, $\gamma_1$, which will be fixed by demanding the asymptotically flat
limit (24). That means that the only independent
parameters are just  $\phi_h$ and $r_h$. Note also that only
the choice $\sigma=+1$ in (60) leads to solutions which have the
desired behaviour (\ref{thirteen}) for the dilaton field at infinity.
Plots involving
the dilaton field $\phi(r)$, for three different allowed values of
the solution parameter $\phi_h$, are given in Figure 1. The metric functions
$e^{\Lambda(r)}$, $e^{\Gamma(r)}$ as well as the three components $T^t_t$,
$T^r_r$ and $T^\theta_\theta$ of the energy-momentum tensor for $r_h=1$
are presented in Figures 2 and 3 respectively.
\pr
As we said before the asymptotic solution near the horizon is
characterized by only two independent parameters, $\phi_h$ and $r_h$.
However, the independent parameters that characterize
the solution near infinity (23)-(25) are three, $M$, $D$ and
$\phi_{\infty}$. From this, we can infer that a relation must hold between the
above parameters in order to be able to classify our solution as a
two parameter family of black hole solutions. After some manipulation,
the set of equations (\ref{sixteen})-(\ref{nineteen}) can
be rearranged to yield the identity
\be
\frac{d}{dr}\left(r^2 e^{(\Gamma-\Lambda)/2}(\Gamma'-\phi')-
\frac{\alpha' e^\phi}{g^2}
e^{(\Gamma-\Lambda)/2} [(1-e^{-\Lambda}) (\phi'-\Gamma')+
e^{-\Lambda} r \phi' \Gamma']\right)=0
\ee
Integrating this relation over the interval $(r_h,r)$ we obtain
the expression
\be
 2 M -D=\sqrt{\gamma_1 \lambda_1} (r_h^2 +\frac{\alpha' e^{\phi_h}}
 {g^2})
\ee
This equation is simply a connection between the set of parameters
describing the solution near the horizon and the set $M$ and $D$.
The rhs of this relation clearly indicates that the existing dependence
of the dilaton charge on the mass does not take the simple form of an
equality encountered in EYMD regular solutions of ref.\cite{donets}.
In order to find the relation between $M$ and $D$ we follow refs.\cite{kanti}
and \cite{mignemi} and take into account the ${\cal O}(\alpha^{'2})$
expression of the dilaton charge in the limit $r \rightarrow \infty$
\bea
\phi(r)&=&\phi_\infty + \frac{D}{r} +... \nn \\ [3mm]
&=&\phi_{\infty} + \left(\frac{e^{\phi_\infty}}{2 M}
 \frac{\alpha'}{g^2} + \frac{73 e^{2 \phi_\infty}}{60 (2 M)^3}
 \frac {\alpha^{'2}}{g^4}\right) \frac{1}{r} +...
\label{dm}
\eea
This relation can be checked numerically.
The result is shown in Figure 4. Any deviations from this relation
are due to higher
order terms which turn out to be small.
\pr
The above relation (\ref{dm}) implies that the dilaton hair
of the black hole solution, discussed in this section,  is a kind
of `secondary hair', in the terminology of ref. \cite{coleman}.
This hair is generated because the basic fields  (gravitons) of
the theory associated
with the primary hair (mass) act as sources for the non-trivial
dilaton
configurations outside the horizon of the black hole.

\section{Additional Solutions}
\pr
A second class of solutions can be obtained if we allow
$\phi'(r)$ to be infinite at some finite value $r_s$ of the coordinate
$r$. This choice, as we shall argue below, is not incompatible with the
finiteness of the energy-momentum tensor.
This is due to the fact that the Gauss-Bonnet
term does not have a definite signature. These solutions have
the same asymptotic characterization in terms of $\phi_{\infty}$,
$M$ and $D$ as the black hole. Near $r \simeq r_s$ one obtains
\bea
e^{-\Lambda(r)}&=& \lambda_1 (r-r_s) +... \nn \\ [3mm]
\Gamma'(r)&=& \frac{\gamma_1}{\sqrt{r-r_s}}+...\\ [3mm]
\phi(r)&=& \phi_s + \phi'_s \sqrt{r-r_s} +...\nn
\label{sol1}
\eea
To lowest order, the equations of motion yield the constraints
\be
\frac{\alpha'}{4 g^2} e^{\phi_s}\phi'_s \gamma_1 = \frac{1}{\lambda_1}
+ \frac{\phi_s^{'2} r_s^2}{16}
\ee
and
\be
 a \gamma_1^2 + b \gamma_1 +c = 0
\ee
with
\bea
 a&=&16 e^{2 \phi_s}(2 e^{\phi_s} \phi^{'2}_s+ 8 r_s + \phi_s^{'2} r_s^2)
 \nn \\ [3mm]
 b&=& 4 e^{\phi_s}(e^{2 \phi_s}\phi_s^{'3} -4 e^{\phi_s} \phi_s^{'3} r_s^2
 -16 \phi'_s r_s^3 -2 \phi_s^{'3} r_s^4) \nn \\ [3mm]
 c&=& -8 e^{2 \phi_s} \phi_s^{'2} r_s -e^{2 \phi_s} \phi_s^{'4} r_s^2 +
 2 e^{\phi_s} \phi_s^{'4} r_s^4 + 8 \phi_s^{'2} r_s^5 + \phi_s^{'6}
\eea
These apparently singular solutions comprise a three-parameter
family.
The behaviour of the dilaton field and the metric components is shown in
Figures 5 and 6 respectively. As we can see, these solutions can
not be classified as
black hole solutions since the metric component $g_{tt}$ does not exhibit
any singular behaviour~: $e^\Gamma \rightarrow constant$
when $r \rightarrow r_s$. Only $g_{rr}$ takes on an infinite value
when $r_s$ is approached.
\pr
In order to determine whether the spacetime geometry is really
singular at $r_s$, the scalar
curvature $R$ as well as the ``curvature invariant"
$I=R^{\mu \nu \rho \sigma}R_{\mu \nu \rho\sigma}$ were calculated.
It turns out that both of
the above quantities do not exhibit any singular
behaviour at $r_s$ which implies that the pathology of the metric is due
to a pathology of the coordinate system and not of the spacetime geometry
itself. Moreover, this guarantees the finiteness of the action (it can be
easily checked that the Gauss-Bonnet combination is also finite which is
consistent with the field redefinition ambiguity arguments
given in the introduction).
It is a simple exercise to verify that the
components of the energy-momentum tensor are
also finite. They are shown in
Figure 7.  Unfortunately, at present, we are not in a position to
discuss the nature of the solutions for $r \le r_s$, and hence
the only safe conclusion to be made from the above analysis
concerns the absence of a {\it naked singularity}.
\pr
It is interesting to mention the existence of another class of solutions
which are regular in the metric, do not possess
any horizon, but the dilaton
becomes infinite at $r \simeq 0$. Near the origin, these solutions are
\bea
e^{\Lambda(r)}&=&1+\lambda_1 e^{-4 \gamma_1/r} \nn \\ [3mm]
\Gamma'(r)&=& \gamma_1 e^{-\phi_1/r} \\ [3mm]
\phi(r)&=& \frac{\phi_1}{r} + ...\nn
\label{sol2}
\eea
This is also a three parameter family of solutions. The far asymptotic
behaviour is still given by (23)-(25), and it is again characterized
by the parameters $\phi_{\infty}$, $M$ and $D$. These solutions
appear to have no curvature singularities ($R (r\simeq0)\simeq 0$), but the
components of the energy-momentum tensor are infinite at $r \simeq 0$.
\pr
Note that our black hole solution appears to be a boundary surface
in the phase space between solutions (66) and (70).
This means that, if $\phi_0$, $\phi_0'$, $\Gamma_0$ and $\Gamma'_0$ are
the values of the fields for the black hole solution at
$r=r_0>>1$, then integration of the system (\ref{thone})-(\ref{thtwo})
starting from $r_0$ with $\phi'(r_0)>\phi'_0$ leads to the
solution (66), whilst the case $\phi'(r_0)<\phi'_0$ leads to the
solution (70).

\section{Conclusions and Outlook}
\pr
In this paper
we have dealt with solutions of the coupled dilaton-graviton
system in four dimensions, in the presence of higher-curvature
terms in the Gauss-Bonnet combination. We have
demonstrated the existence of black hole solutions for
this system, characterized
by non-trivial scalar (dilaton) hair.
This hair is of `secondary' type, in the
sense that it
is not accompanied by the
presence of any new quantity that characterizes the black hole.
Indeed, it was shown above that
the dilaton charge
is not an independent quantity, but it can be
expressed in terms of the mass of the black hole.
It should be
stressed, however,
that irrespectively of the precise type of hair
the set of solutions examined in this work
bypasses the conditions of the no-hair theorem~\cite{bek}.
Thus, our solutions may be viewed as
demonstrating that there is plenty
of room in the gravitational
structure of Superstring Theory to allow for
physically sensible situations that are not covered by the
theorem as stated. Although our results were derived in the
framework of the ${\cal O}(\alpha')$ effective superstring action,
they are non-perturbative in nature and they will persist at
least in situations of moderate curvatures.
\pr
In addition to the black hole solutions, we were able to
find two other families of solutions,
one of which had the interesting feature of having
finite energy density. At present, the physical
significance of the solutions is not fully clear to us.
We hope to be able to return and study
these structures
in the near future.
\pr
There are many features of the solutions which we did not address
in this work, one of which is their stability under either linear
time-dependent perturbations of the graviton-dilaton multiplet,
or under generic
perturbations
(beyond linearity).
Such an analysis has been
performed for the Einstein-Yang-Mills-Higgs
system~\cite{mw}, and one could think of extending it to incorporate
higher-curvature gravity theories. A stability analysis, when
completed, will prove essential in understanding better
the physical significance of the black hole
solutions found
in this work. This is of particular interest
due to
the connection of the solutions
with superstring theory. We hope to return
to these issues in a future publication.

\pr
\noindent {\Large {\bf Acknowledgements}}
\pr
P.K., N.E.M., and J.R. would like to thank the Theory Division at CERN
for the hospitality during the final stages of this work. Two of us
(K.T and P.K.) acknowledge travelling support by the EEC Human Capital
and Mobility Network ``Flavourdynamics" (CHRX-CT93-0132). In addition, P.K.
acknowledges financial support for travelling to CERN by the Greek
Ministry of Technology. E.W wishes to thank EPSRC (U.K.) for a
Research Studentship.

\newpage
\begin{figure}
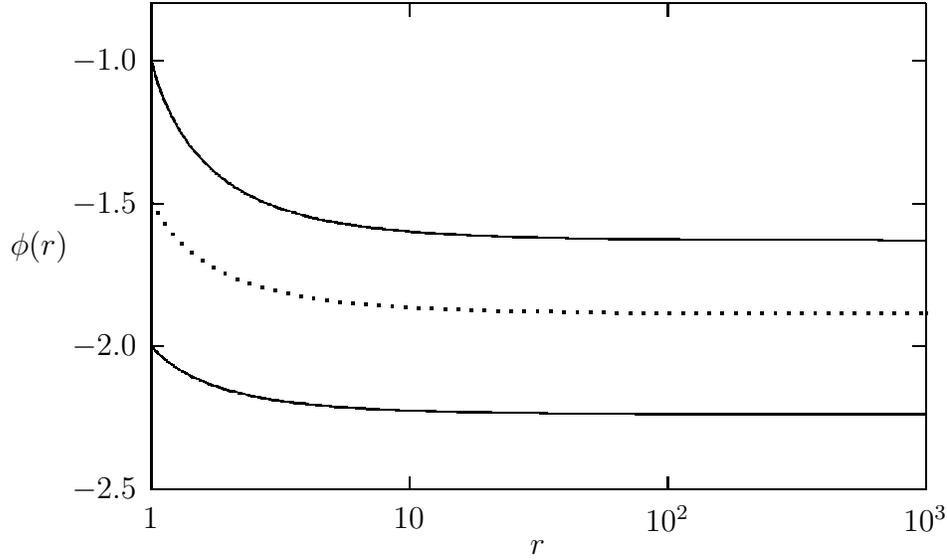

\begin{center}
%
\setlength{\unitlength}{0.240900pt}
\ifx\plotpoint\undefined\newsavebox{\plotpoint}\fi
\sbox{\plotpoint}{\rule[-0.200pt]{0.400pt}{0.400pt}}%


\caption{Dilaton field for $r_h=1$ black hole. Each curve
corresponds to a different solution characterized by a
different initial value of $\phi_h$.}
\end{center}
\end{figure}
\begin{figure}
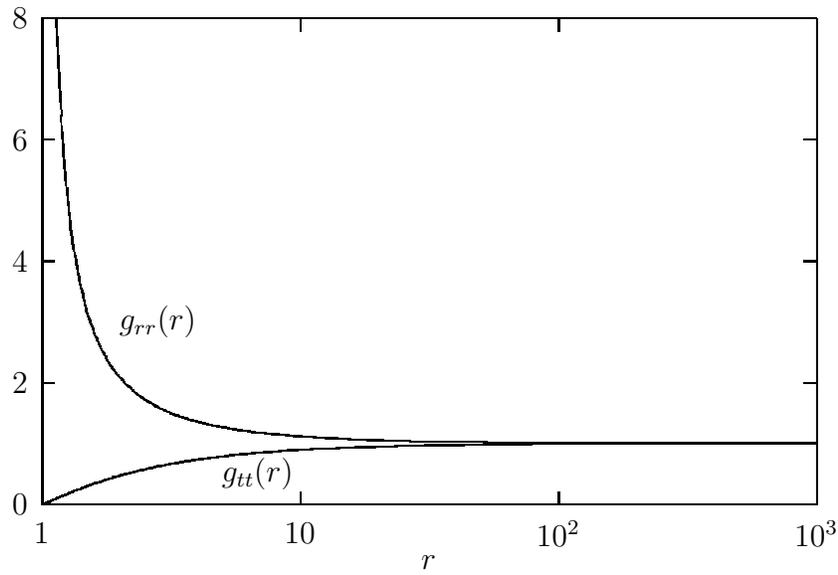

\begin{center}
\setlength{\unitlength}{0.240900pt}
\ifx\plotpoint\undefined\newsavebox{\plotpoint}\fi
\sbox{\plotpoint}{\rule[-0.200pt]{0.400pt}{0.400pt}}%
%

\caption{Metric components $g_{tt}$ and $g_{rr}$ for $r_h=1$
black hole.}
\end{center}
\end{figure}
\begin{figure}
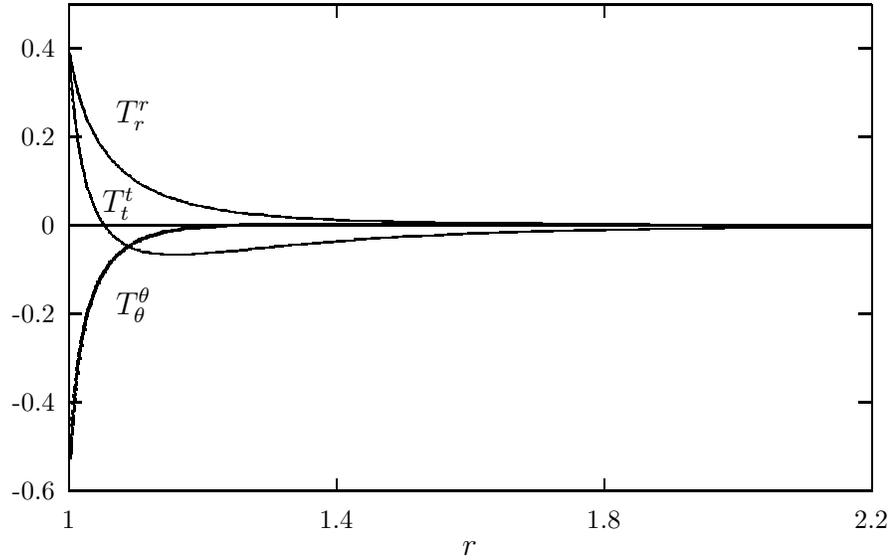

\begin{center}
\setlength{\unitlength}{0.240900pt}
\ifx\plotpoint\undefined\newsavebox{\plotpoint}\fi
%

\caption{Components of the energy-momemtum tensor for $r_h=1$
black hole.}
\end{center}
\end{figure}
\begin{figure}
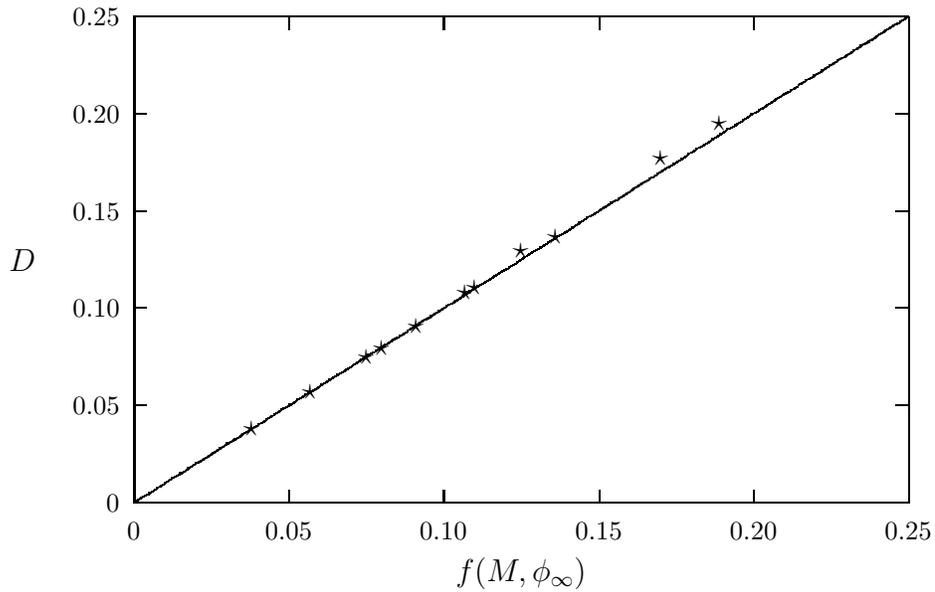

\begin{center}
\setlength{\unitlength}{0.240900pt}
\ifx\plotpoint\undefined\newsavebox{\plotpoint}\fi
\sbox{\plotpoint}{\rule[-0.200pt]{0.400pt}{0.400pt}}%
%

\caption{Dependence of the dilaton charge $D$ on $M$ and
$\phi_\infty$ for the $r_h=1$ black hole. The function
$f(M,\phi_\infty)$ stands for the coefficient of $1/r$ in
the eq.(\ref{dm}).}
\end{center}
\end{figure}
\begin{figure}
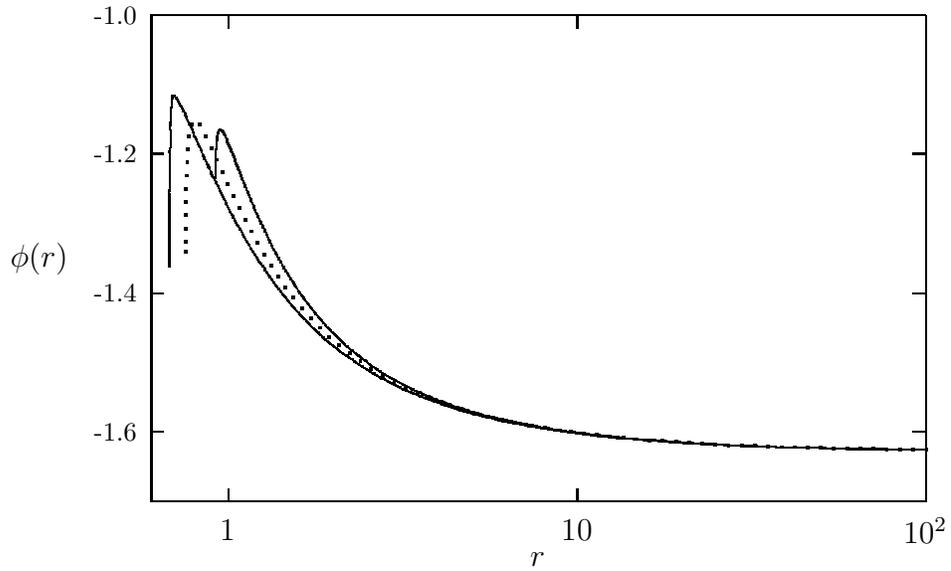

\begin{center}
%
\setlength{\unitlength}{0.240900pt}
\ifx\plotpoint\undefined\newsavebox{\plotpoint}\fi
\sbox{\plotpoint}{\rule[-0.200pt]{0.400pt}{0.400pt}}%
%
\caption{Dilaton field for the singular solution $(66)$. Each curve
corresponds to a different solution characterized by a
different value of $r_s$ ($r_s= 0.92,0.75,0.62$).}
\end{center}
\end{figure}
\begin{figure}
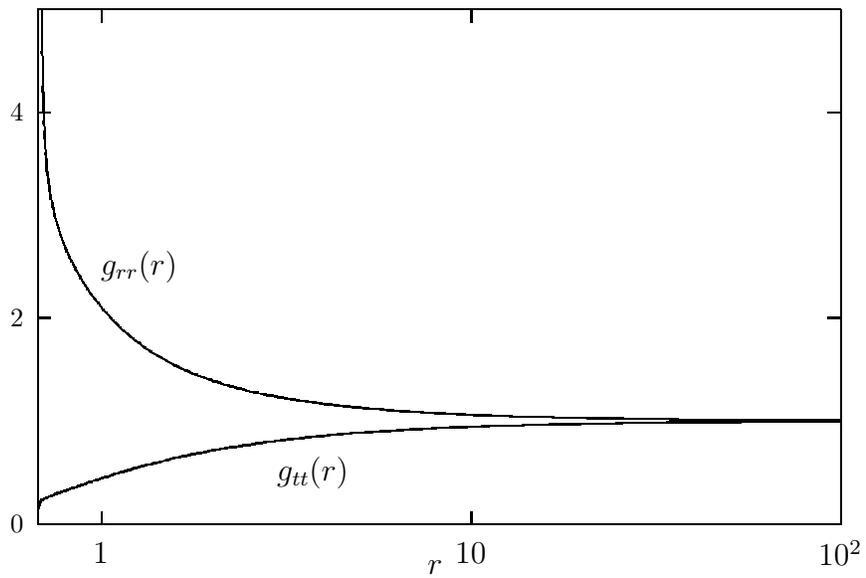

\begin{center}
%
\setlength{\unitlength}{0.240900pt}
\ifx\plotpoint\undefined\newsavebox{\plotpoint}\fi
%
\caption{Metric components for the $r_s=0.68$ singular solution $(66)$.}
\end{center}
\end{figure}
\begin{figure}
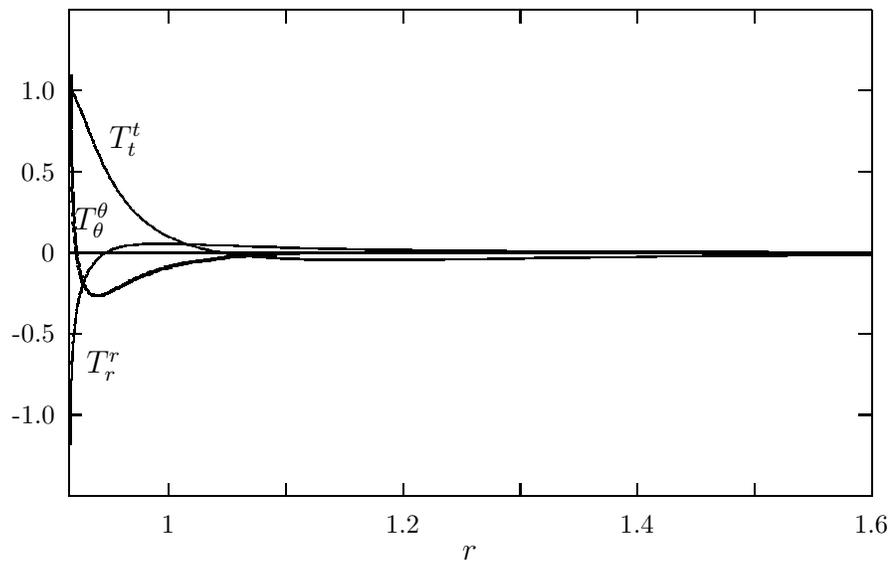

\begin{center}
%
\setlength{\unitlength}{0.240900pt}
\ifx\plotpoint\undefined\newsavebox{\plotpoint}\fi
\sbox{\plotpoint}{\rule[-0.200pt]{0.400pt}{0.400pt}}%
%
\caption{Components of the energy-momentum tensor for the
$r_s=0.92$ singular solution (66).}
\end{center}
\end{figure}
\end{document}